\documentclass[numberedappendix,apj,twocolumn]{emulateapj}

\usepackage[bookmarks,bookmarksnumbered,colorlinks=true, citecolor=blue, linkcolor=black,breaklinks]{hyperref}
\usepackage{epstopdf}

\newcommand\Tstrut{\rule{0pt}{2.6ex}}         

\shorttitle{An HST Grism Redshift at $z=11$}
\shortauthors{Oesch et al.}

\submitted{Draft version \today}

\begin{document}

\title{A Remarkably Luminous Galaxy at z\,=\,11.1 Measured with Hubble Space Telescope Grism Spectroscopy}

\author{P. A. Oesch\altaffilmark{1,2}, 
G. Brammer\altaffilmark{3}, 
P. G. van Dokkum\altaffilmark{1,2},
G. D. Illingworth\altaffilmark{4}, 
R. J. Bouwens\altaffilmark{5}, 
I. Labb\'{e}\altaffilmark{5}, 
M. Franx\altaffilmark{5}, 
I. Momcheva\altaffilmark{2,3},
M. L. N. Ashby\altaffilmark{6}, 
G. G. Fazio\altaffilmark{6}, 
V. Gonzalez\altaffilmark{7,8},
B. Holden\altaffilmark{4},
D. Magee\altaffilmark{4},
R. E. Skelton\altaffilmark{9},
R. Smit\altaffilmark{10},
L. R. Spitler\altaffilmark{11,12},
M. Trenti\altaffilmark{13},
S. P. Willner\altaffilmark{6}}

\altaffiltext{1}{Yale Center for Astronomy and Astrophysics, Yale University, New Haven, CT 06511, USA}
\altaffiltext{2}{Astronomy Department, Yale University, New Haven, CT 06511, USA}
\altaffiltext{3}{Space Telescope Science Institute, 3700 San Martin Drive, Baltimore, MD 21218, USA}
\altaffiltext{4}{UCO/Lick Observatory, University of California, Santa Cruz, 1156 High St, Santa Cruz, CA 95064, USA}
\altaffiltext{5}{Leiden Observatory, Leiden University, NL-2300 RA Leiden, The Netherlands}
\altaffiltext{6}{Harvard-Smithsonian Center for Astrophysics, Cambridge, MA 02138, USA}
\altaffiltext{7}{Departamento de Astronomia, Universidad de Chile, Casilla 36-D, Santiago, Chile}
\altaffiltext{8}{Centro de Astrofisica y Tecnologias Afines (CATA), Camino del Observatorio 1515, Las Condes, Santiago, Chile}
\altaffiltext{9}{South African Astronomical Observatory, P.O. Box 9, Observatory 7935, South Africa}
\altaffiltext{10}{Department of Physics, Durham University, South Road, Durham DH1 3LE, UK}
\altaffiltext{11}{Department of Physics and Astronomy, Faculty of Sciences, Macquarie University, Sydney, NSW 2109, Australia}
\altaffiltext{12}{Australian Astronomical Observatory, P.O. Box 915, North Ryde, NSW 1670, Australia}
\altaffiltext{13}{School of Physics, University of Melbourne, Parkville 3010, VIC, Australia}

\begin{abstract}
We present $Hubble$ WFC3/IR slitless grism spectra of a remarkably
bright $z\gtrsim10$ galaxy candidate, GN-z11, identified initially from CANDELS/GOODS-N imaging
data. A significant spectroscopic continuum break is detected at $\lambda=1.47\pm0.01~\mu$m. The new
grism data, combined with the photometric data, rule out all plausible lower
redshift solutions for this source. The only viable solution is that
this continuum break is the Ly$\alpha$ break redshifted to ${z_\mathrm{grism}=11.09^{+0.08}_{-0.12}}$, 
just $\sim$400 Myr after the Big Bang. This observation extends the current spectroscopic frontier by 150
Myr to well before the Planck (instantaneous) cosmic reionization peak at ${z\sim8.8}$, 
demonstrating that galaxy build-up was well underway early in the
reionization epoch at $z > 10$. GN-z11 is remarkably and unexpectedly luminous for a galaxy
at such an early time: its UV luminosity is 3${\times}$ larger than ${L_*}$ measured at ${z\sim6-8}$.  
The $Spitzer$ IRAC detections up to 4.5~$\mu$m of this galaxy are consistent with a stellar mass of ${\sim10^{9}~M_\odot}$. This spectroscopic redshift
measurement suggests that the \textit{James Webb Space Telescope} ($JWST$) will be able to similarly and easily confirm such
sources at $z>10$ and characterize their physical properties through detailed spectroscopy. Furthermore,
WFIRST, with its wide-field near-IR imaging, would find large numbers of
similar galaxies and contribute greatly to $JWST$'s spectroscopy, if it is launched early enough to overlap with $JWST$. 
\end{abstract}


\vspace*{0.4truecm}

\section{Introduction}

The first billion years are a crucial epoch in cosmic history. This is when the first stars and galaxies formed and the universe underwent a major phase transition from a neutral to an ionized state. Our understanding of galaxies in this early phase of the universe has been revolutionized over the last few years thanks to the very sensitive WFC3/IR camera onboard the Hubble Space Telescope (\textit{HST}) in combination with ultra-deep \textit{Spitzer}/IRAC imaging. WFC3/IR has pushed the observational horizon of galaxies to the beginning of the cosmic reionization epoch at $z\sim9-11$, less than 500 Myr from the Big Bang.
Several large extragalactic surveys have now resulted in the identification of a large sample of more than 800 galaxies at $z\sim7-8$ \citep{Bouwens15aLF,McLure13,Finkelstein14,Bradley13,Schmidt14} and even a small sample of $z\sim9-11$ candidates \citep{Oesch13,Oesch14,Oesch15b,Ellis13,Zheng12,Coe13,Zitrin14,Bouwens15,McLeod15,Ishigaki15,Infante15,Kawamata15,Calvi16}.

Spectroscopic confirmations of very high-redshift candidates remain limited, however. 
The primary spectral feature accessible from the ground for these sources, the Ly$\alpha$ line, is likely attenuated by the surrounding neutral hydrogen for all $z>6$ galaxies \citep{Schenker12,Treu13,Pentericci14}. Therefore, despite the large number of candidates from \textit{HST} imaging, only a handful of galaxies in the epoch of reionization have confirmed redshifts to date \citep{Vanzella11,Ono12,Shibuya12,Finkelstein13,Oesch15,Roberts-Borsani15,Zitrin15}. 

Given the low success rate of Ly$\alpha$ searches, a viable alternative approach is to search for a spectroscopic confirmation of the UV continuum spectral break \citep[see e.g.][]{DowHygelund05,Malhotra05,Vanzella09,Rhoads13,Watson15,Pirzkal15}. This  break is expected owing to the near-complete absorption of UV photons shortward of Ly$\alpha$ by neutral hydrogen in the early universe. For the brightest known $z\gtrsim7$ candidates ($H\lesssim26$ AB mag), the continuum flux is within reach of the powerful WFC3/IR grism spectrometer given the low near-infrared background at the orbit of \textit{HST}. Several surveys have thus been undertaken or are ongoing to search for continuum breaks and very weak Ly$\alpha$ lines with deep WFC3/IR grism spectra \citep[e.g.][]{Treu15,Schmidt15,Pirzkal15}.

Our team recently discovered a small sample of $z\gtrsim9$ candidate galaxies bright enough to test this approach at the highest accessible redshifts with HST \citep{Oesch14}. In our analysis of the public CANDELS data over the GOODS fields, we identified six relatively bright ($H_{160}=26.0-26.8$ mag) galaxies with best-fit photometric redshifts $z=9.2-10.2$. 
These sources more than doubled the number of known galaxy candidates at 500 Myr. Remarkably they were $\sim$10--20 times more luminous than any prior candidate. The question thus arose whether these bright galaxies really are at $z>9$ or whether they are part of a previously unknown population at lower redshifts. 
While the photometric data strongly indicate that the candidates are very high-redshift galaxies, one could not completely rule out extreme emission line galaxies at lower redshift.

This paper presents the results of a 12 orbit WFC3/IR grism spectroscopy program (GO-13871, PI:Oesch) which targeted the intrinsically most luminous $z\gtrsim10$ galaxy candidate among our previous sample. This paper is organized as follows: Section \ref{sec:data} summarizes the grism spectroscopy and ancillary imaging data before we present the resulting spectrum, 
which provides strong evidence for a continuum Lyman-break at $z=11.1$ (see Section \ref{sec:results}). We end with a short discussion of our findings in Section \ref{sec:discussion}.

Throughout this paper, we adopt $\Omega_M=0.3, \Omega_\Lambda=0.7, H_0=70$ kms$^{-1}$Mpc$^{-1}$, i.e., $h=0.7$, largely consistent with the most recent measurements from Planck \citep{Planck2015}. Magnitudes are given in the AB system \citep{Oke83}, and we will refer to the \textit{HST} filters F435W, F606W, F814W, F105W, F125W, F140W, F160W as $B_{435}$, $V_{606}$, $I_{814}$, $Y_{105}$, $J_{125}$, $JH_{140}$, $H_{160}$, respectively.

\begin{deluxetable}{ll}
\tablecaption{Photometry of GN-z11}
\tablecolumns{2}
\tablewidth{0.8\linewidth}

\tablehead{Filter & Flux Density [nJy]}

\startdata
  $B_{435}$ & $7 \pm 9$    \\
 $V_{606}$ & $2 \pm 7$  \\ 
 $i_{775}$ & $5 \pm 10$  \\
 $I_{814}$ & $3 \pm 7$  \\
 $z_{850}$ & $17 \pm 11$  \\
 $Y_{105}$ & $-7 \pm 9$  \\
 $J_{125}$ & $11 \pm 8$  \\
 $JH_{140}$ & $64 \pm 13$\tablenotemark{$\dagger$} \\  
 $H_{160}$ & $152 \pm 10$ \\
 $K$ & $137 \pm 67$   \\ 
 $\mathrm{IRAC}~ 3.6\,\mu \mathrm{m}$ & $139 \pm 21$\tablenotemark{$\dagger$} \\ 
 $\mathrm{IRAC}~ 4.5\,\mu \mathrm{m}$ & $144 \pm 27$\tablenotemark{$\dagger$} 
\enddata

\tablenotetext{$\dagger$}{Note that the earlier JH$_{140}$ data in which GN-z11 was first detected was less deep and gave a flux measurement of $104\pm47$ nJy. For other previous photometry measurements see Table 3 in \citet[][]{Oesch14}. }
\label{tab:photometry}
\end{deluxetable}

\section{Target Selection and Data}
\label{sec:data}

\subsection{Target Selection}
Galaxy samples at $z\sim9-10$ are now being assembled based on deep \textit{HST} imaging data \citep{Zheng12,Oesch13,Ellis13,Coe13,Bouwens15,McLeod15}. Such distant sources can be identified based on a continuum break in the $J_{125}$ band, i.e., at around 1.2~$\mu$m. In a recent analysis of the WFC3/IR imaging data from the public CANDELS survey \citep{Koekemoer11,Grogin11}, we identified six surprisingly bright sources ($H_{160}=26.0-26.8$ mag) with photometric redshifts $z=9.2-10.2$ in the two GOODS fields \citep{Oesch14}.

Interestingly, a large fraction of these luminous $z>9$ galaxy candidates were detected individually in the rest-frame optical with \textit{Spitzer}/IRAC \citep{Fazio04} data from the S-CANDELS survey \citep{Ashby15}. This provides a sampling of the spectral energy distribution (SED) of these sources from the rest-frame UV to the rest-frame optical. The  strong breaks measured in their $J_{125}-H_{160}$ colors, the complete non-detections in the optical images, and the blue colors from \textit{HST} to IRAC significantly limit the contamination by known low-redshift SEDs and together point to a true high-redshift nature for all these sources. Stellar contamination could be ruled out based on colors and sizes. Furthermore, no evidence for contamination by an active galactic nucleus was found based on variability over a 1 year timescale and X-ray upper limits \citep{Oesch14}. 
Nevertheless, spectroscopic confirmation is clearly required to verify the high-redshift solutions for these targets.

The brightness of these $z\sim9-10$ galaxy candidates puts them within reach of WFC3/IR grism continuum spectroscopy, opening up the possibility of obtaining a grism redshift based on a continuum break. 
The brightest of these candidates, GN-z11, also had the highest photometric redshift and was thus identified as the target for follow-up spectroscopy for our \textit{HST} program GO-13871 (PI: Oesch). 

GN-z11 lies in the CANDELS DEEP area in GOODS-North at (RA,DEC$)=($12:36:25.46, +62:14:31.4) and has $H_{160} = 26.0\pm0.1$. It was previously introduced as GN-z10-1 \citep{Oesch14} or as source 20253 in the 3D-HST photometric catalog \citep{Skelton14}. GN-z11 also has the highest S/N in both IRAC bands (3.6 and 4.5\,$\mu$m) among the sample of $z\sim9-10$ galaxies from \citet{Oesch14}. The $H_{160}$ profile of GN-z11 shows clear asymmetry with its isophotal area extending over nearly 0\farcs6. A stellar source can therefore be excluded.

The depth of the data used to originally identify this candidate was 27.8 mag in $H_{160}$, 27.0 mag in IRAC channel 1, and 26.7 mag in channel 2 \citep[all 5$\sigma$; see][]{Oesch14}. 
The photometry of the target source is listed in Table \ref{tab:photometry}. Compared to our discovery paper \citep{Oesch14}, this includes new measurements in the two IRAC 3.6 and 4.5 $\mu$m filters as well as in the WFC3/IR $JH_{140}$ band: deeper data are now available in $JH_{140}$ as part of our grism pre-imaging (see next section), and the IRAC photometry was updated based on a new, independent reduction of all available IRAC imaging data in the Spitzer archive by our team \citep[similar to][]{Labbe15}.

\begin{figure}[htb]
    \centering
    \includegraphics[width=0.96\linewidth]{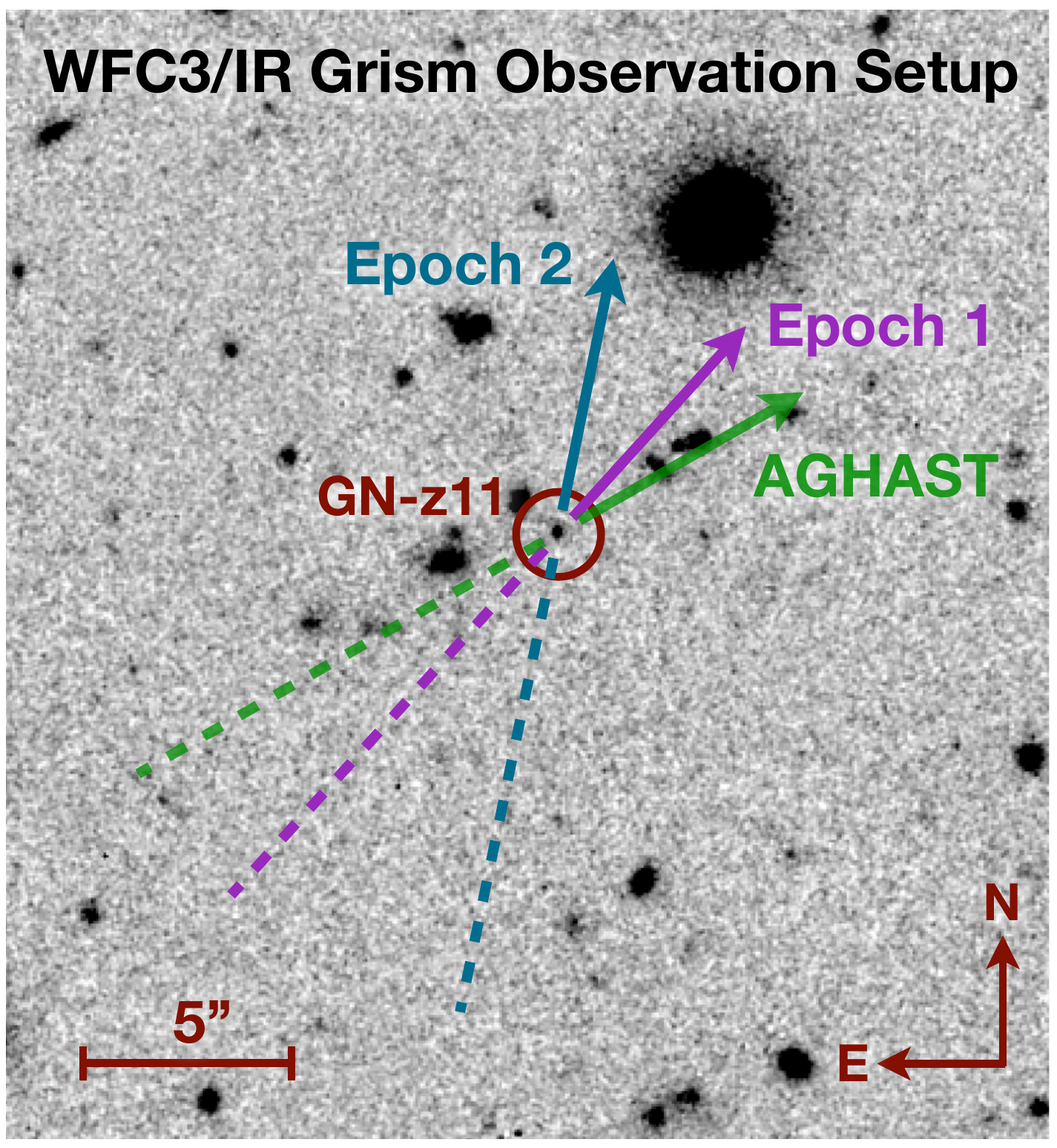}
    \caption{ 
CANDELS $H$-band image around the location of our target source GN-z11. The arrows and dashed lines indicate the direction along which sources are dispersed in the slitless grism spectra for our two individual epochs (magenta and blue) and for the pre-existing AGHAST data (green). The latter are significantly contaminated by bright neighbors along the dispersion direction of GN-z11 (see Fig \ref{fig:ContamModel}).
 }
 \label{fig:GrismOrient}
 \end{figure}

\begin{figure}[htb]
    \centering
    \includegraphics[width=\linewidth]{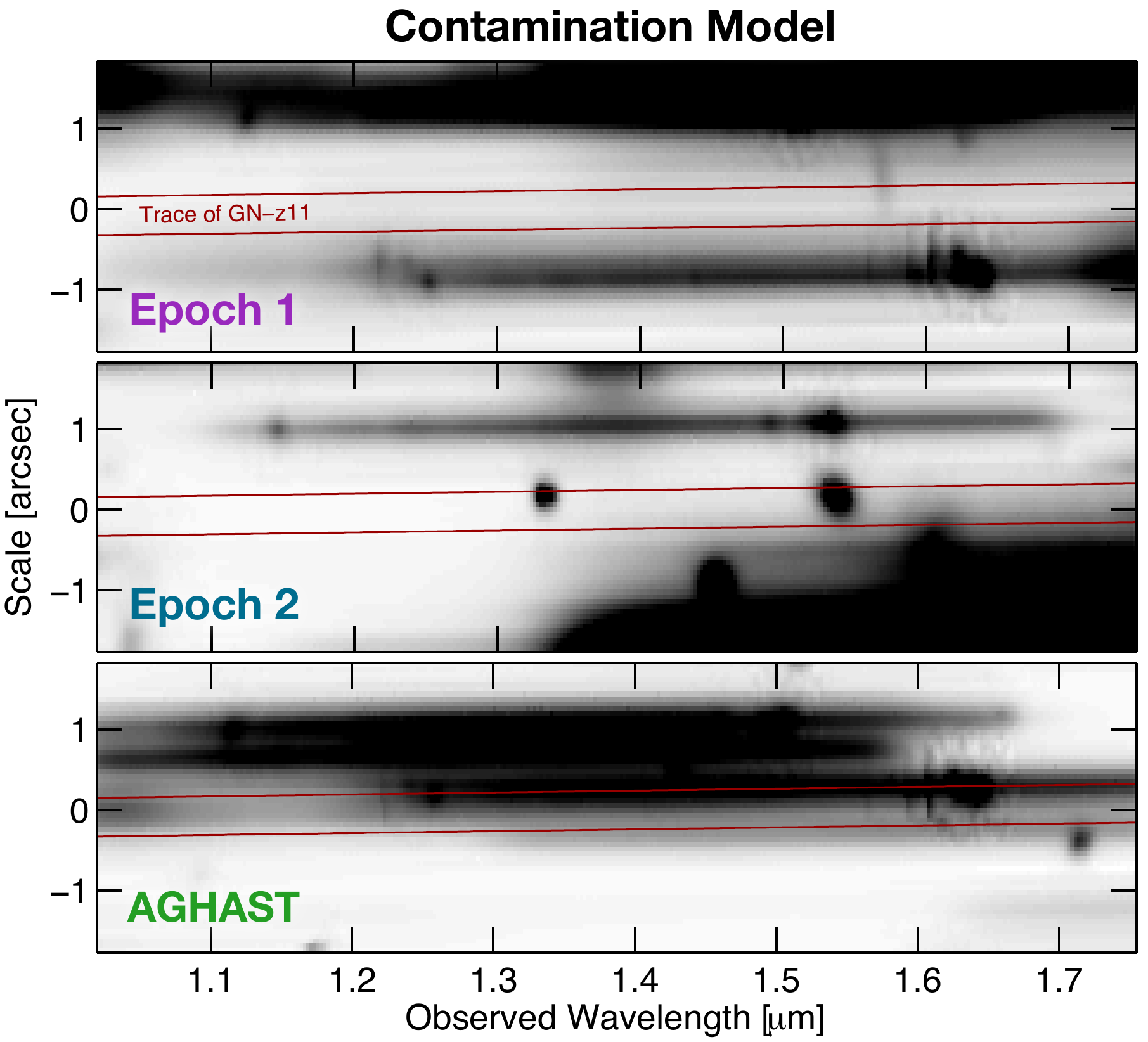}
    \caption{ 
Our model of the contaminating flux from neighboring sources in the slitless grism spectra around the trace of our source (i.e., the line along which we expect its flux; indicated by red lines). From top to bottom, the panels show the final model contamination in our spectra in epochs 1 and 2 and in the pre-existing AGHAST spectra. Note that the contamination model includes emission lines for the neighboring sources as calibrated from our two-epoch data. The high contaminating flux in the AGHAST data makes these spectra inadequate for studying GN-z11. Our orientations were chosen based on extensive simulations to minimize such contamination from neighbors. However, some zeroth order flux in epoch 2 could not be avoided while at the same time making the observations schedulable with \textit{HST} in cycle 22.
 }
 \label{fig:ContamModel}
 \end{figure}

\newpage

\subsection{Slitless Grism Data}
The primary data analyzed in this paper are 12 orbit deep G141 grism spectra from our \textit{HST} program GO-13871 (PI: Oesch). These spectra were taken at two different orients in two epochs of six orbits each on 2015 February 11 and April 3 (see Figure \ref{fig:GrismOrient}). The data acquisition and observation planning followed the successful 3D-HST grism program \citep{Brammer12,Momcheva15}. 
Together with each G141 grism exposure, a short 200~s pre-image with the $JH_{140}$ filter was taken to determine the zeroth order of the grism spectra. The $JH_{140}$ images were placed at the beginning and end of each orbit in order to minimize the impact of variable sky background on the grism exposure due to the bright Earth limb and He 1.083 $\mu$m line emission from the upper atmosphere \citep{BrammerReport14}. A four-point dither pattern was used to improve the sampling of the point-spread function and to overcome cosmetic defects of the detector.

The grism data were reduced using the grism reduction pipeline developed by the 3D-HST team. The main reduction steps are explained in detail by \citet[][]{Momcheva15}. In particular, the flat-fielded and global background-subtracted grism images are interlaced to produce 2D spectra for sources at a spectral sampling of $\sim$23 \AA, i.e., about one quarter of the native resolution of the G141 grism. 
The final 2D spectrum is a weighted stack of the data from individual visits.
In particular, we down-weight pixels which are affected by neighbor-contamination using 
\begin{equation}
w = [ (2*f_{contam})^2+\sigma^2 ] ^{-1}
\label{eq:1}
\end{equation} 

where $f_{contam}$ is the contamination model flux in a particular pixel and $\sigma$ is the per-pixel RMS taken from the WFC3/IR noise model \citep[c.f., \S 3.4.3 from][]{Rajan11}. We have tested that $\sigma$ is accurately characterized as demonstrated by the pixel flux distribution function in the 2D residual spectra (see Appendix Fig \ref{fig:residualHist}).

A local background is subtracted from the 2D frame which is a simple 2nd order polynomial estimated from the contamination-cleaned pixels above and below the trace of the target source.
1D spectra are then computed using optimal extraction on the final 2D frames, weighting the flux by the morphology of the source as measured in the WFC3/IR imaging \citep[][]{Horne86}.

The GOODS-N field, which contains our candidate GN-z11, has previously been covered by 2 times 2-orbit deep grism data from ``A Grism H-Alpha SpecTroscopic survey" (AGHAST; GO-11600; PI: Wiener). 
However, the spectrum of our target GN-z11 is significantly contaminated in the AGHAST observations by several nearby galaxies given their orientation (see Figure \ref{fig:ContamModel}). Furthermore, the observations were severely affected by variable sky background \citep{BrammerReport14}. The final data used in this paper therefore do {\it not} include the AGHAST spectra, which were not optimized for GN-z11 that was discovered 3--4 years after they were taken. We confirmed that including the AGHAST data would not affect our results, however, because of the down-weighting of heavily contaminated pixels in our stacking procedure.

\newpage
\subsection{The Challenge of Cleaning Slitless Grism Spectra}
One major challenge in slitless grism spectroscopy is the systematic contamination of the target spectrum by light from nearby galaxies. Our spectra were therefore taken at two orientations with the grism dispersion offset by 32 degrees and tailored to show the least possible contamination while being schedulable (see Figure \ref{fig:ContamModel}).
The optimal orientations were determined through extensive simulations of the location of all orders from -1\textsuperscript{st} to +3\textsuperscript{rd} of all sources in the field detected in the extensive ancillary imaging datasets. Nevertheless, some contaminating flux can never be avoided and needs to be modeled. Following the same techniques as developed for the 3D-HST survey, this was done by fitting all the neighboring galaxies' SEDs  (i.e., the EAZY template fits to the Skelton et al. photometric catalog) and using their morphologies to create  2D spectral models for all of them, which were then subtracted from the 2D spectrum of our target GN-z11 spectrum.

\begin{figure}[tb]
    \centering
    \includegraphics[width=0.98\linewidth]{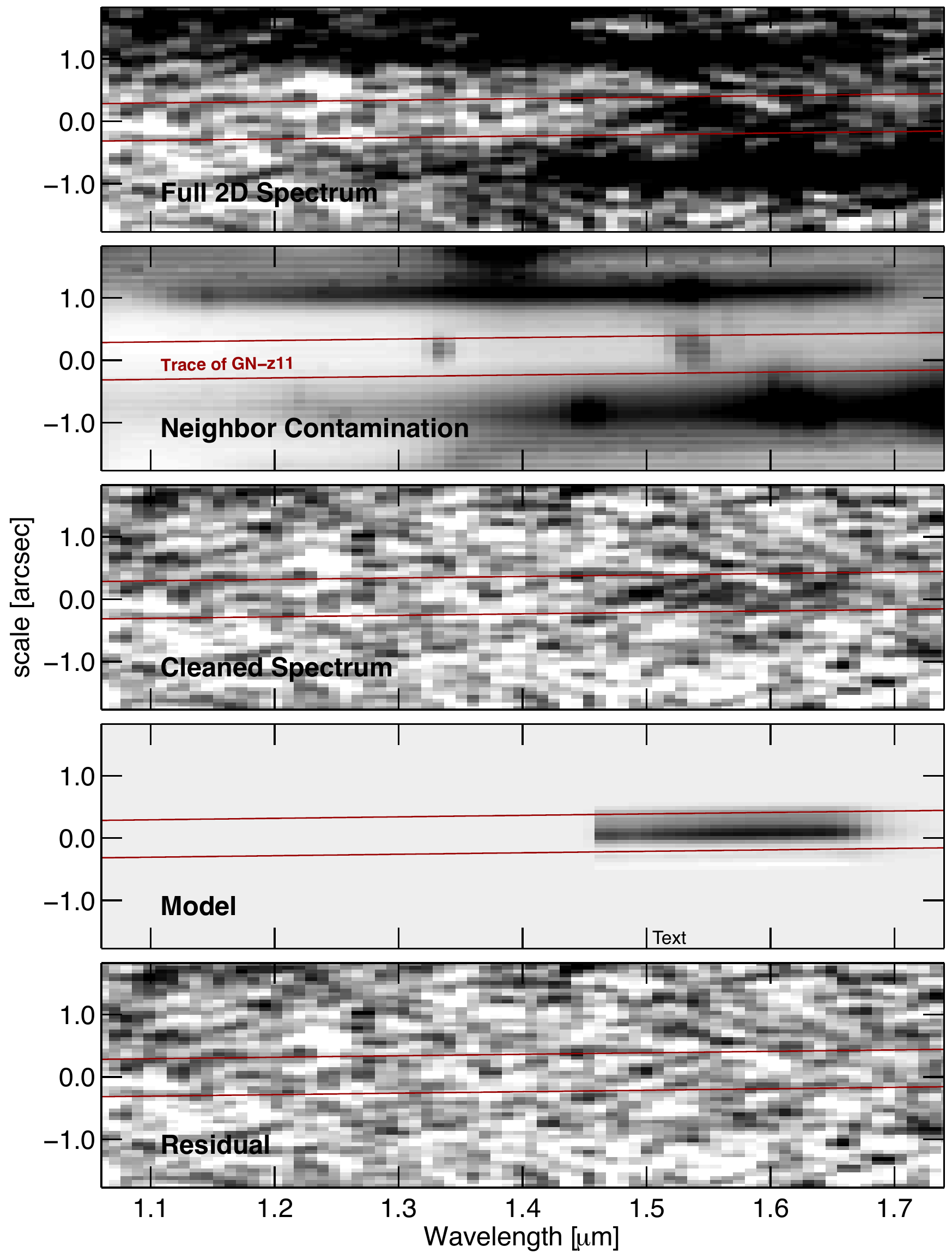}
   \vspace{-3pt}    
    \caption{ 2D grism data of GN-z11.
The five panels show from top to bottom (1) the original 2D spectrum from a stack of all our G141 grism data, (2) the modeled contaminating flux from neighboring sources, (3) the cleaned 2D spectrum of GN-z11, (4) the model of a $z=11.09$ source with the same morphology and H-band magnitude as GN-z11, and (5) the residual spectrum after subtracting the $z=11.09$ continuum model. 
The observed grism flux is completely consistent with the model flux, as can be seen from the clean residual. The observed spectrum also falls off at $\sim1.65~\mu$m, exactly as expected based on the drop in the G141 grism sensitivity providing further strong support that the observed flux is indeed the continuum of GN-z11.
The spatial direction extends over 3.6 arcsec, and the two red lines indicate the trace of GN-z11. 
 }
 \label{fig:2Dgrism}
 \end{figure}

Having access to deep grism spectra at multiple independent orients also allows us to further refine the 2D spectral model of neighboring galaxies by including actually measured emission line fluxes. We thus extracted spectra of all neighboring sources at both orientations and fit their spectra as outlined by \citet{Momcheva15}. The resulting best fits were then used directly as contamination models. This produces significantly cleaner 2D spectra compared to assuming simple continuum emission templates for contamination modeling. The resulting contamination of all neighboring sources in our data as well as in the AGHAST spectra are shown in Figure \ref{fig:ContamModel}. The quality of this quantitative contamination model is also discussed in the next sections and shown in Figure \ref{fig:2Dgrism}.

\begin{figure*}[tb]
    \centering
    \includegraphics[width=0.95\linewidth]{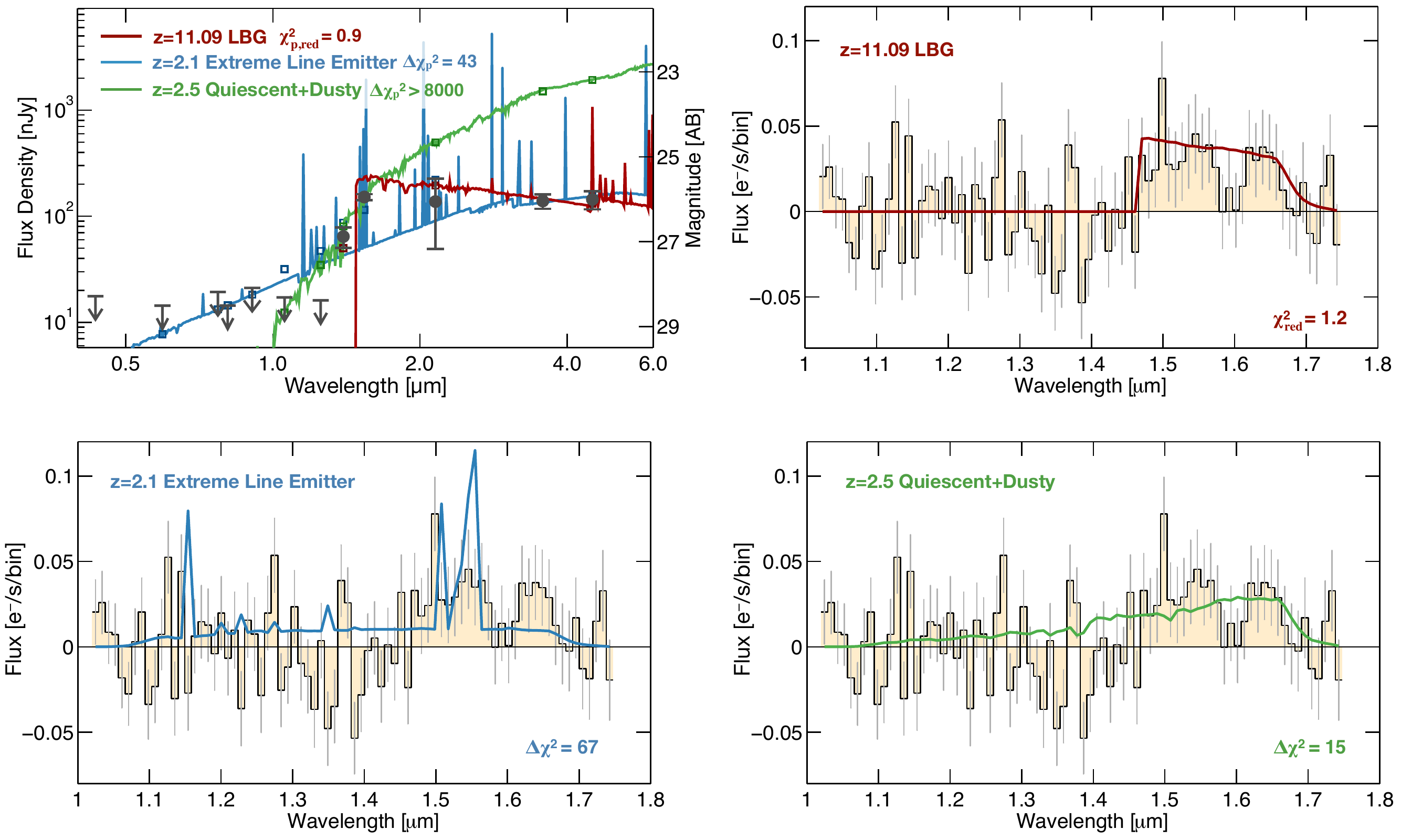}
   \vspace{-3pt}    
    \caption{ 
The new 12-orbit deep grism spectra in combination with the photometry of GN-z11 exclude lower redshift solutions. The main contaminants for high-redshift galaxy selections are sources with extreme emission lines or with very strong 4000 \AA\ breaks. The top left  panel shows the photometry together with three example SEDs for the possible nature of GN-z11 (dark red: a $z=11.09$ star-forming galaxy, blue: an extreme line emitter at $z=2.1$, green: a dusty+quiescent galaxy at $z=2.5$). The last one is only shown for illustration purposes as it can be clearly excluded based on the longer wavelength photometry ($\Delta\chi_p^2>8000$ relative to the best fit SED model).
The remaining panels compare the observed 1D spectrum with the expected grism fluxes for the same three cases. The best-fit to the grism data is provided by the high-redshift LBG template which interprets the observed break as a Ly$\alpha$ break. This solution has a reduced $\chi^2=1.2$. The other two cases can be excluded based on the difference in $\chi^2$ in the grism spectra as well as from the photometry ($\chi_p^2$).
 }
 \label{fig:ContamSpectra}
 \end{figure*}

\begin{figure*}[tb]
  \begin{center}
  \includegraphics[width=.65\linewidth]{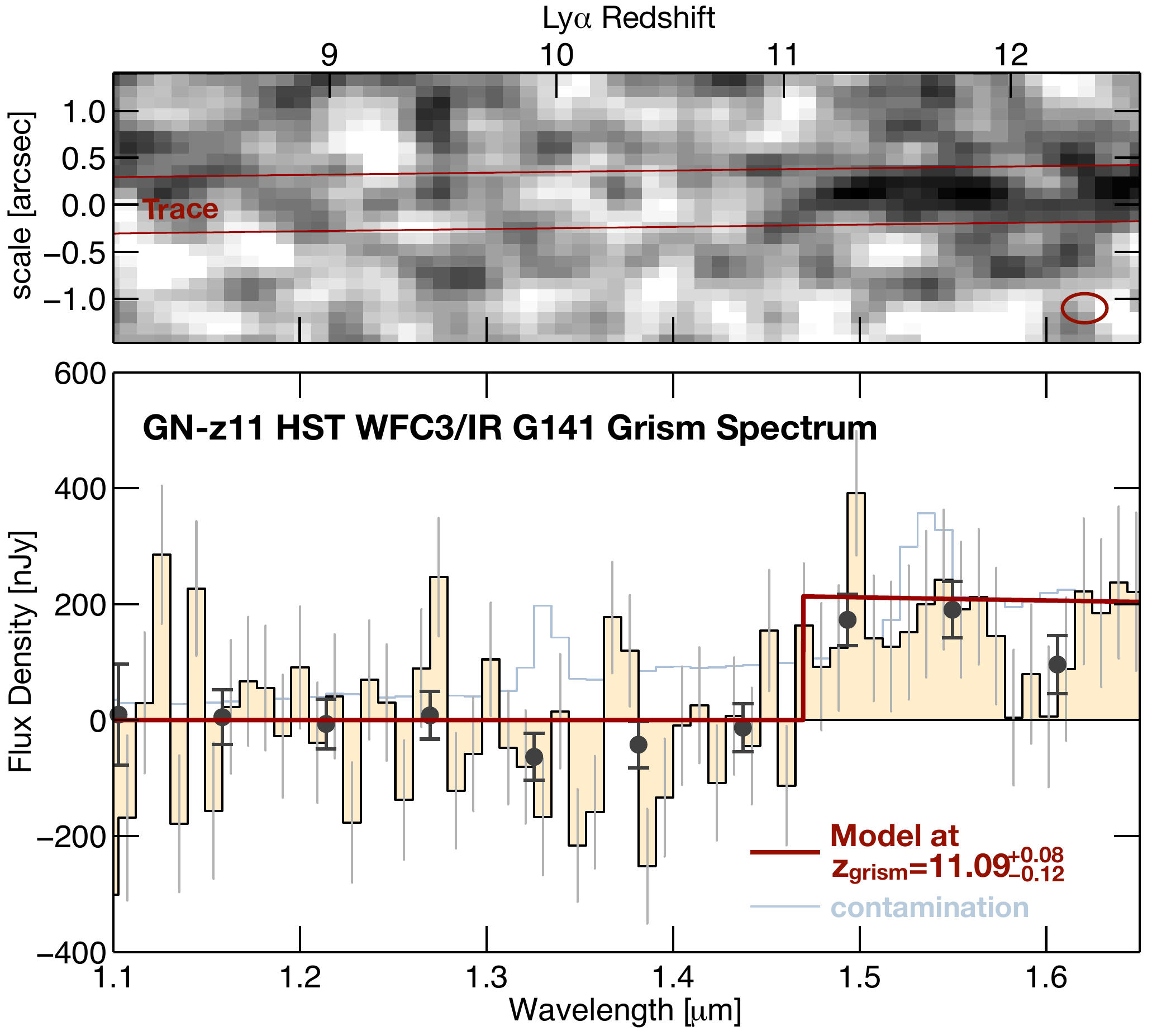}
  \end{center}
    \caption{ {Grism Spectrum of GN-z11.} The top panel shows the (negative) 2D  spectrum from the stack of our cycle 22 data (12 orbits) with the trace outlined by the dark red lines. For clarity the 2D spectrum was smoothed by a Gaussian indicated by the ellipse in the lower right corner.
The bottom panel is the un-smoothed 1D flux density using an optimal extraction rebinned to one resolution element of the G141 grism (93~\AA). The black dots show the same further binned to 560~\AA, while the blue line shows the contamination level that was
subtracted from the original object spectrum. We identify a continuum break in the spectrum at $\lambda=1.47\pm0.01~\mu$m.
    The continuum flux at $\lambda>1.47~\mu m$ is detected at $\sim1-1.5\sigma$ per resolution element and at 3.8$\sigma$ per 560~\AA\ bin.
After excluding lower redshift solutions (see text and Fig \ref{fig:ContamSpectra}), the best-fit grism redshift is   $z_\mathrm{grism}=11.09^{+0.08}_{-0.12}$.    
The red line reflects the Ly$\alpha$ break at this redshift, normalized to the measured H-band flux of GN-z11. The agreement is excellent.
    The fact that we only detect significant flux along the trace of our target source, which is also consistent with the measured H-band magnitude, is strong evidence that we have indeed detected the continuum of GN-z11 rather than any residual contamination.    
   }
   \label{fig:currentGrism}
\end{figure*}

\section{Continuum Detection}
\label{sec:results}

The final stacked 2D grism spectrum is shown in the top panel of Figure \ref{fig:2Dgrism}. Clearly, contamination is significant outside of the expected trace of GN-z11. After subtracting our detailed contamination model, however, we obtain a clean 2D frame (middle panel of Fig \ref{fig:2Dgrism}). This shows clear flux exactly along the dispersion location of GN-z11. After rebinning to a spectral resolution of 93~\AA\ (the native resolution element of the grism) this flux detection is $\sim1-1.5\sigma$ per resolution element longward of 1.47~$\mu$m and consistent with zero flux shortward of that. The total spectral flux averaged over 1.47-1.65~$\mu$m represents a clear 5.5$\sigma$ detection. This is fully consistent with the prediction from the exposure time calculator for an $H=26$ mag source in a 12-orbit exposure. 

The extracted 1D spectrum along the trace of GN-z11 is shown in Figures \ref{fig:ContamSpectra} and \ref{fig:currentGrism}. 
These highlight the detection of a continuum break with a flux ratio of $f(\lambda<1.47)/f(\lambda>1.47) < 0.32$ at $2\sigma$ when averaged over 560 \AA\ wide spectral bins.

A flux decrement is seen around $\sim1.6$~$\mu$m, which is caused by negative flux values in one of our two visits slightly above the peak of the trace of GN-z11. However, this dip is consistent with Gaussian deviates from the noise model. We also tested that the detected continuum flux is still present when adopting different stacking and extraction procedures (see appendix \ref{app:grism}). In particular, we confirmed the spectral break in a simple median stack of the individual continuum-subtracted 2D grism observations of our 6 individual visits. Additionally, we confirmed that the continuum break is seen only along the trace of our spectrum by creating simple 1D-extractions above and below the trace of our target source. 
Finally, we confirmed that a break is seen in both epochs separately (see Fig \ref{fig:SpecByEpoch}).

In the following sections, we discuss the possible redshift solutions for GN-z11 based on the combined constraints of the new grism continuum detection as well as the pre-existing photometry.

\subsection{The Best-fit Solution: A $z\sim11$ Galaxy}
Based on our previous photometric redshift measurement for GN-z11 ($z_\mathrm{phot}=10.2$), we expected to detect a continuum break at $1.36\pm0.05~\mu$m. This can clearly be ruled out. However, the grism data are consistent with an even higher redshift solution.
Interpreting the observed break as the 1216~\AA\ break, which is expected for high-redshift galaxies based on absorption from the neutral inter-galactic hydrogen along the line of sight, we obtain a very good fit to the spectrum with a reduced $\chi^2=1.2$ (see Figs \ref{fig:ContamSpectra} and \ref{fig:currentGrism}). The best fit redshift is $z_\mathrm{grism}=11.09^{+0.08}_{-0.12}$, corresponding to a cosmic time of only $\sim$400 Myr after the Big Bang.
The grism redshift and its uncertainty are derived from an MCMC fit to the 2D spectrum which also includes the morphological information of the source as well as the photometry, adopting identical techniques as used for the 3D-HST survey redshifts \citep[see][]{Momcheva15}.

This high-redshift solution also reproduces the expected count rate based on an $H=26$ mag continuum source, as well as the overall morphology of the 2D grism spectrum. This is demonstrated in Figure \ref{fig:2Dgrism}, where we show the original data, the neighbor subtracted 2D spectrum as well as the residual after subtracting out the $z=11.09$ model with the correct $H$-band magnitude. Note the drop of the flux longward of 1.65~$\mu$m due to the reduced sensitivity of the grism. This is an important constraint, because it shows that the detected flux originates from the source itself and is not due to residual neighbor contamination. 

Figure \ref{fig:residualHist} (in the appendix) further shows that the pixel distribution of the residual 2D frame is in excellent agreement with the expectations from pure Gaussian noise. This demonstrates that our contamination subtraction and flux uncertainty estimates were derived appropriately and that the resulting values are accurate.

Despite the difference from the previous photometric redshift estimate, the measured grism redshift is consistent with the photometry of this source (see Figure \ref{fig:SEDfits}). While our previous photometric redshift estimate was $z_\mathrm{phot}=10.2\pm0.4$, the redshift likelihood function contained a significant tail to $z>11$. The updated and deeper $JH_{140}$ photometry subsequently resulted in a shift of the peak by $\Delta_z=0.2$ to a higher redshift.
The $z=11.09$ solution is within 1$\sigma$ of the now better measured $JH_{140}-H_{160}$ color, which is the main driver for the photometric redshift estimate, as shown in Figure \ref{fig:SEDfits}. The grism data significantly tighten the redshift likelihood function (bottom panel Figure \ref{fig:SEDfits}) in addition to excluding lower redshift solutions.

In the next two sections we also show that we can safely exclude all plausible lower redshift solutions.
The new grism redshift confirms that this source lies well beyond the peak epoch of cosmic reionization  \citep[$z_\mathrm{reion}=8.8$;][]{Planck2015} and makes it the most distant known galaxy. This includes sources with photometric redshift measurements, apart from a highly debated source in the HUDF/XDF field, which likely lies at $z\sim2$ but has a potential $z\sim12$ solution \citep[see][]{Bouwens11a,Bouwens13b,Ellis13,Brammer13}.

\subsection{Excluding a Lower-Redshift Strong Line Emitter}
The principal goal of our grism program was to unequivocally exclude a lower redshift solution for the source GN-z11. While GN-z11 shows a very strong continuum break with $J_{125}-H_{160}>2.4$ ($2\sigma$), without a spectrum, we could not exclude contamination by a source with very extreme emission lines with line ratios reproducing a seemingly flat continuum longward of 1.4 $\mu$m \citep{Oesch14}. 

The previous AGHAST spectra already provided some evidence against strong emission line contamination \citep{Oesch14}, and we also obtained Keck/MOSFIRE spectroscopy to further strengthen this conclusion (see appendix). However, the additional 12 orbits of G141 grism data now conclusively rule out that GN-z11 is such a lower redshift source. Assuming that all the $H$-band flux came from one emission line, we would have detected this line at $>10\sigma$. Even when assuming a more realistic case where the emission line flux is distributed over a combination of lines (e.g., H$\beta + [$\ion{O}{3}$]$), we can confidently invalidate such a solution. The lower left panel in Figure \ref{fig:ContamSpectra} compares the measured grism spectrum with that expected for the best-fit lower redshift solution we had previously identified \citep{Oesch14}. A strong line emitter SED is clearly inconsistent with the data. Apart from the emission lines, which we do not detect, this model also predicts weak continuum flux across the whole wavelength range. At $<1.47$~$\mu$m, this is higher than the observed mean, while at $>1.47$~$\mu$m the expected flux is too low compared to the observations.
Overall the likelihood of a $z\sim2$ extreme emission line SED based on our grism data is less than 10$^{-6}$ and can be ruled out.

Note that in very similar grism observations for a source triply imaged by a CLASH foreground cluster, emission line contamination could also be excluded \citep{Pirzkal15}. We thus have no indication currently that any of the recent $z\sim9-11$ galaxy candidates identified with \textit{HST} is a lower redshift strong emission line contaminant \citep[but see, e.g.,][for a possible $z\sim12$ candidate]{Brammer13}.

\subsection{Excluding a Lower-Redshift Dusty or Quiescent Galaxy}
Another potential source of contamination for very high redshift galaxy samples are dusty $z\sim2-3$ sources with strong 4000 \AA\ or Balmer breaks \citep{Oesch12a,Hayes12}. However, the fact that the IRAC data for GN-z11 show that it has a very blue continuum longward of 1.6~$\mu$m, together with the very red color in the WFC3/IR photometry, already rules out such a solution (see SED plot in Figure \ref{fig:ContamSpectra}). Nevertheless, we additionally explore what constraints the grism spectrum alone can set on such a solution. 

The expected flux for such a red galaxy increases gradually across the wavelength range covered by the G141 grism, unlike what we observe in the data (lower right in Fig \ref{fig:ContamSpectra}). Compared to our best-fit solution (see next section) we measure a $\Delta\chi^2 = 15$ when comparing the data with the expected grism flux. Apart from the extremely large discrepancy with the IRAC photometry, we can thus exclude this solution at 98.9\% confidence based on the spectrum alone.

Similar conclusions can be drawn  from the break strength alone \citep[see e.g.][]{Spinrad98}.
Assuming that the observed break at 1.47 $\mu$m corresponds to 4000 \AA\ at $z=2.7$, a galaxy with a maximally old spectral energy distribution (single burst at $z=15$) would show a flux ratio of $(1-f_\nu^{short}/f_\nu^{long})<0.63$ when averaged over 560 \AA\ bins. This is based on simple \citet{Bruzual03} models without any dust. As mentioned earlier, the observed spectrum has a break of $(1-f_\nu^{short}/f_\nu^{long})>0.68$ at $2\sigma$, thus indicating again that we can marginally rule out a 4000 \AA\ break based on the spectrum alone without even including the photometric constraints.

\begin{figure}[tb]
    \centering
    \includegraphics[width=\linewidth]{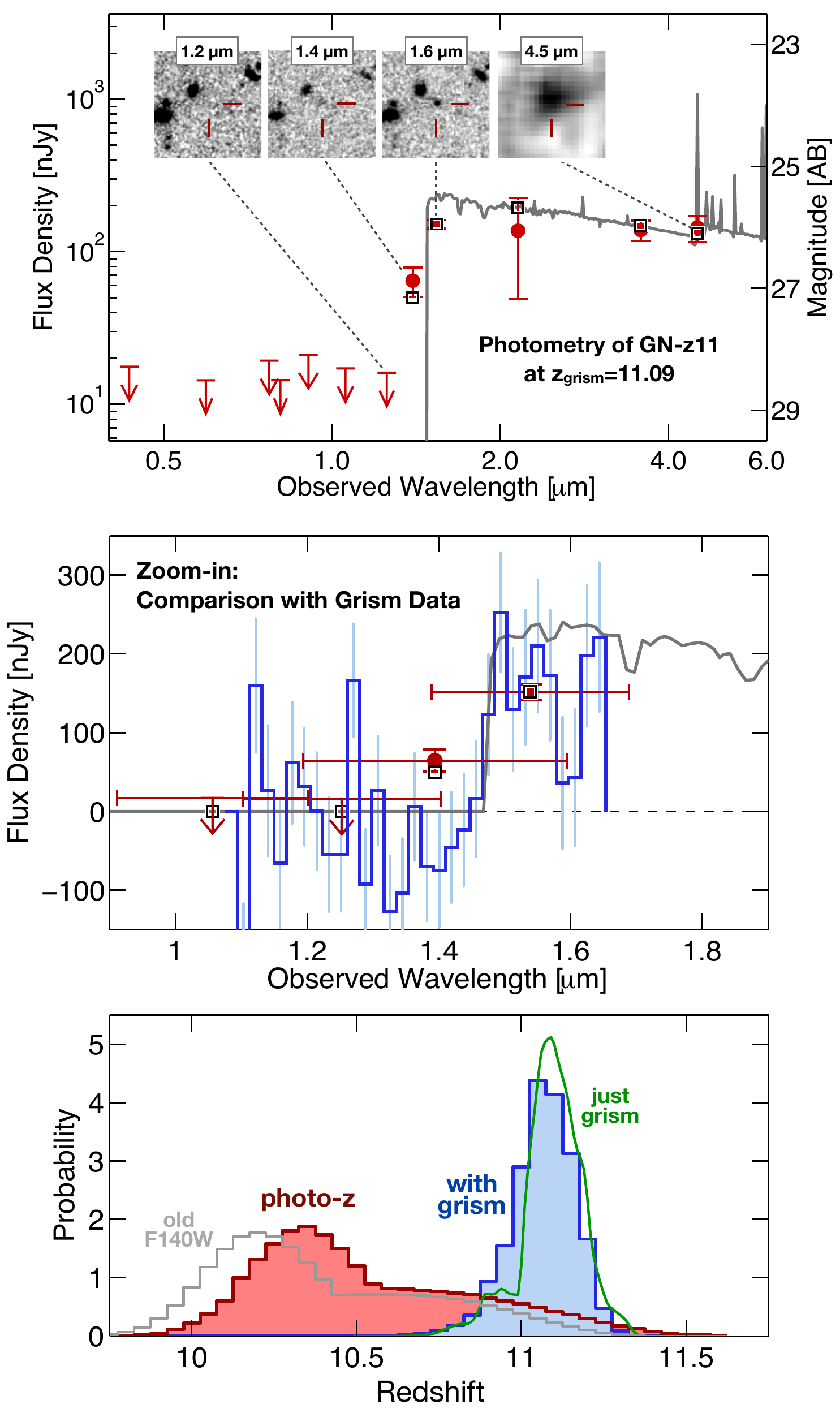}
   \vspace{-3pt}    
    \caption{ 
\textit{Top -- } The photometry (red) and best-fit spectral energy distribution (SED; gray) of GN-z11 at the measured grism redshift of $z=11.09$. Upper limits correspond to 1$\sigma$ non-detections. The black squares correspond to the flux measurements of the best-fit SED. 
Inset negative images of 6\arcsec$\times$6\arcsec\ show the \textit{HST} $J_{125}$, $JH_{140}$, and $H_{160}$ bands as well as the neighbor-cleaned IRAC 4.5~\micron\ image.  
GN-z11 is robustly detected in all bands longward of 1.4~\micron, resulting in accurate constraints on its physical parameters. The photometry is consistent with a galaxy stellar mass of $\log M/M_\odot\sim9.0$ with no or very little dust extinction and a young average stellar age.
\textit{ Middle -- } A zoom-in around the wavelength range probed by the G141 grism. The rebinned grism data are shown by the blue line with errorbars. The grism flux is consistent with the photometry (red points) and the best-fit SED at $z=11.09$ (gray line). The red horizontal errorbars represent the wavelength coverage of the different HST filters, indicating that the break of GN-z11 lies  within the $H_{160}$ band.
\textit{ Bottom -- } The redshift probability distribution functions, $p(z)$, when fitting only to the broad-band photometry (red) or when including both the photometry and the grism in the fits (blue). The photometric $p(z)$ peaks at significantly lower redshift, but contains an extended tail to $z>11$. The addition of the grism data significantly tightens the $p(z)$ resulting in uncertainties of $\Delta z\simeq0.1$. The fits that include the old, shallower $JH_{140}$ photometry are shown in
gray, while those that use the grism data alone, without any photometric
constraints, are shown in green.
 }
 \label{fig:SEDfits}
 \end{figure}

\section{Discussion}
\label{sec:discussion}

\subsection{Physical Properties of GN-z11}

Despite being the most distant known galaxy, GN-z11 is relatively bright and reliably detected in both IRAC 3.6 and 4.5~$\mu$m bands from the S-CANDELS survey \citep{Ashby15}. This provides a sampling of its rest-frame UV spectral energy distribution and even partially covers the rest-frame optical wavelengths in the IRAC 4.5~$\mu$m band (see Figure \ref{fig:SEDfits}).

The photometry of GN-z11 is consistent with a spectral energy distribution (SED) of $\log M/M_\odot\sim9$ using standard templates \citep[][see appendix]{Bruzual03}.
The UV continuum is relatively blue with a UV spectral slope $\beta=-2.5\pm0.2$ as derived from a powerlaw fit to the $H_{160}$, $K$, and [3.6] fluxes only, indicating very little dust extinction \citep[see also][]{Wilkins15}. Together with the absence of a strong Balmer break, this is consistent with a young stellar age of this galaxy. The best fit age is only 40 Myr ($<110$ Myr at $1\sigma$). GN-z11 thus formed its stars relatively rapidly. The inferred star-formation rate is 24$\pm10$ M$_\odot$/yr. All the inferred physical parameters for GN-z11 are summarized in Table \ref{tab:specsummary}. Overall, our results show that galaxy build-up was well underway at $\sim400$ Myr after the Big Bang.

\begin{deluxetable}{ll}
\tablecaption{Summary of Measurements for GN-z11}
\tablecolumns{2}
\tablewidth{0.8\linewidth}

\startdata
\hline
$\mathrm{R.A.}$ &  $12:36:25.46$ \Tstrut  \\
$\mathrm{Dec.}$ &  $+62:14:31.4$ \\
$\mathrm{Redshift}~ z_\mathrm{grism}$  &  $11.09^{+0.08}_{-0.12}$\tablenotemark{$a$}  \\
$\mathrm{UV~ Luminosity}~ M_{UV}$  & $-22.1\pm0.2$ \\
$\mathrm{Half-Light~ Radius}$\tablenotemark{$b$}   & $0.6\pm0.3~\mathrm{kpc}$\\
$\log M_{gal}/M_\odot$ \tablenotemark{$c$}   &  $9.0\pm0.4$ \\
$\log \mathrm{age/yr}$ \tablenotemark{$c$}   & $7.6\pm0.4$  \\
$\mathrm{SFR}$ & $24\pm10 ~M_\odot~\mathrm{yr}^{-1}$ \\    
$\mathrm{A}_\mathrm{UV}$  & $<0.2~ \mathrm{mag}$ \\
$\mathrm{UV~ slope}~ \beta ~~(f_\lambda\propto \lambda^\beta)$  & $-2.5\pm0.2$\tablenotemark{$d$}   
\enddata

\tablenotetext{$a$}{Age of the Universe at $z=11.09$ using our cosmology: 402 Myr }
\tablenotetext{$b$}{From \citet{Holwerda15}}
\tablenotetext{$c$}{Uncertainties are likely underestimated, since our photometry only partially covers the rest-frame optical for GN-z11}
\tablenotetext{$d$}{See also \citet{Wilkins15}}
\label{tab:specsummary}
\end{deluxetable}

\subsection{The Number Density of Very Bright $z>10$ Galaxies}

The spectrum of GN-z11 indicates that its continuum break lies within the $H_{160}$ filter (which covers $\sim1.4-1.7$~$\mu$m; see Fig \ref{fig:SEDfits}). The rest-frame UV continuum flux of this galaxy is therefore $\sim$0.4 mag brighter than inferred from the $H_{160}$ magnitude. The estimated absolute magnitude is  $M_{UV}=-22.1\pm0.2$, which is roughly a magnitude brighter (i.e., a factor 3$\times$) than the characteristic luminosity of the UV luminosity function at $z\sim7-8$ \citep{Bouwens15aLF,Finkelstein14}.  
With $z_\mathrm{grism}=11.09$, the galaxy GN-z11 is thus surprisingly bright and distant (see Figure \ref{fig:currentGrism}). While one single detection of a galaxy this bright is not very constraining given the large Poissonian uncertainties, it is interesting to estimate how many such galaxies we could have expected based on (1) the currently best estimates of the UV LF at $z>8$ and (2) based on theoretical models and simulations. 

Our target was found in a search of the GOODS fields, which amount to  $\sim160$ arcmin$^2$. However, in a subsequent search of the three remaining CANDELS fields no similar sources were found with likely redshifts at $z\gtrsim10$ \citep{Bouwens15}. We therefore use the full 750 arcmin$^2$ of the CANDELS fields with matching WFC3/IR and ACS imaging for a volume estimate, which amounts to 1.2$\times$10$^6$ Mpc$^3$ (assuming $\Delta z=1$).

Using the simple trends in the Schechter parameters of the UV LFs measured UV at lower redshift ($z\sim4-8$) and extrapolating these to $z=11$, we can get an empirical estimate of the number density of very bright galaxies at $z\sim11$. This amounts to 0.06 \citep{Bouwens15aLF} or 0.002 \citep{Finkelstein14} expected galaxies brighter than $M_\mathrm{UV}=-22.1$ in our survey corresponding to less than 0.3 per surveyed square degree. 
Similarly, recent empirical models \citep{Mashian15,Mason15,Trac15} predict only 0.002$-$0.03 galaxies as bright as GN-z11 in our survey or 0.01$-$0.2 per deg$^2$. All the assumed LF parameters together with the resulting estimates of the number of expected bright galaxies $N_{exp}$ are listed in Table \ref{tab:assumedLFs}.

The above estimates illustrate that our discovery of the unexpectedly luminous galaxy GN-z11 may challenge our current understanding of galaxy build-up at $z>8$.
A possible solution is that the UV LF does not follow a Schechter function form at the very bright end as has been suggested by some authors at $z\sim7$ \citep{Bowler14}, motivated by inefficient feedback in the very early universe. However, current evidence for this is still weak \citep[see discussion in][]{Bouwens15aLF}. Larger area studies will be required in the future \citep[such as the planned WFIRST High Latitude Survey;][]{Spergel15} surveying several square degrees to determine the bright end of the UV LF to resolve this puzzle.

\begin{figure}[tb]
    \centering
        \includegraphics[width=\linewidth]{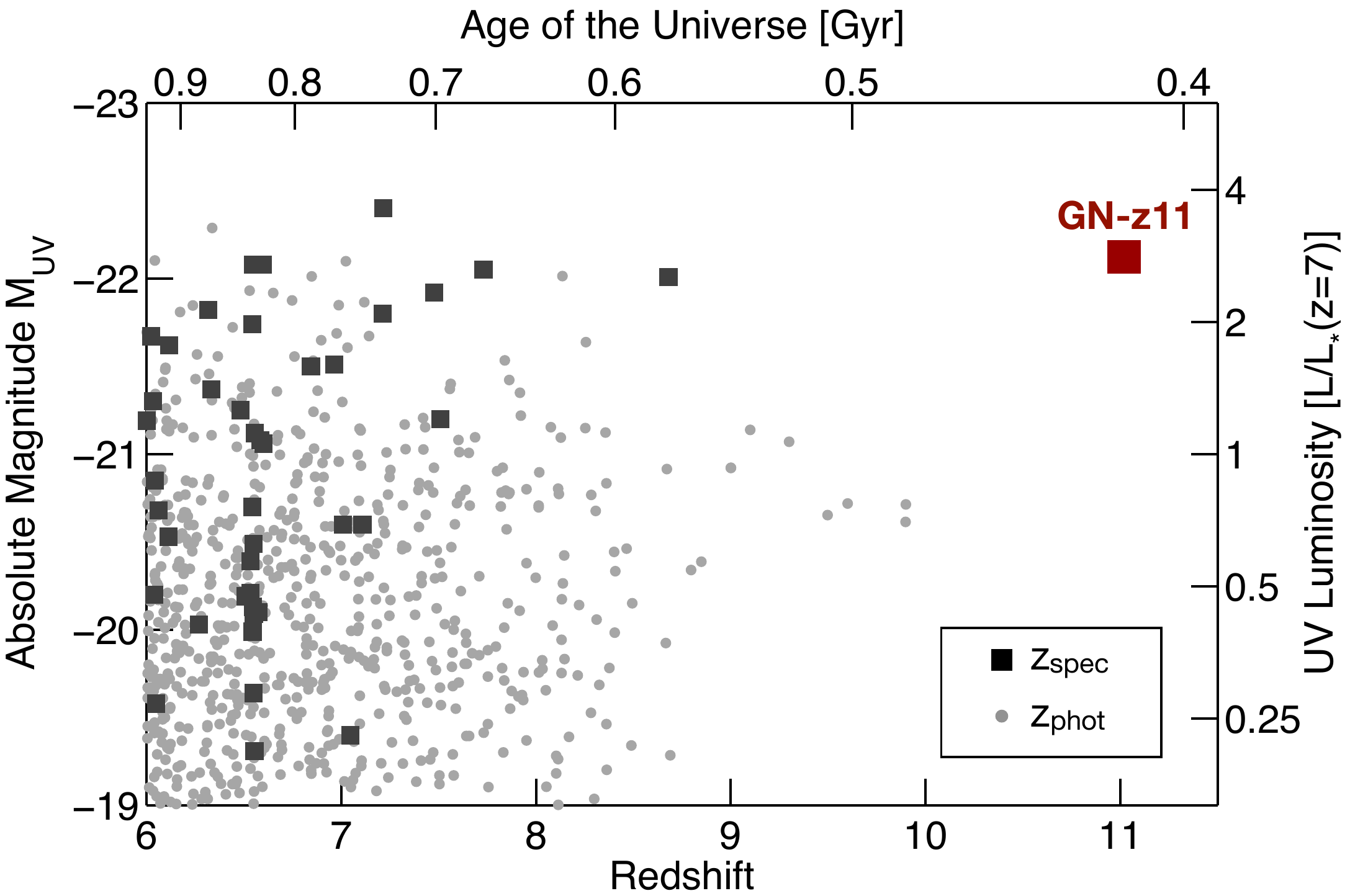}
   \vspace{-3pt}    
    \caption{ 
The redshift and UV luminosities of known high-redshift galaxies from blank field surveys. Dark filled squares correspond to spectroscopically confirmed sources, while small gray dots are photometric redshifts \citep{Bouwens15aLF}. GN-z11 clearly stands out as one of the most luminous currently known galaxies at all redshifts $z>6$ and is by far the most distant measured galaxy with spectroscopy \citep[black squares; see][for a full list of references]{Oesch15}. Wider area surveys with future near-infrared telescopes (such as WFIRST) will be required to determine how common such luminous sources really are at $z>10$.
 }
 \label{fig:ZspecMuv}
 \end{figure}

\begin{deluxetable}{lcccc}
\tablecaption{Assumed LFs for $z\sim10-11$ Number Density Estimates}
\tablecolumns{5}
\tablewidth{\linewidth}

\tablehead{Reference & $\phi*/10^{-5}$ & $M*$ & $\alpha$ & $N_\mathrm{exp}$ \\
& [Mpc$^{-3}$] & [mag] & & $(<-22.1)$ }

\startdata
\citet{Bouwens15aLF} &  1.65 & -20.97 & -2.38  & 0.06 \\ 
\citet{Finkelstein14} & 0.96 & -20.55 & -2.90  & 0.002 \\
\citet{Mashian15} & 0.25 & -21.20 & -2.20  & 0.03 \\ 
\citet{Mason15} & 0.30 & -21.05 & -2.61  & 0.01 \\
\citet{Trac15} & 5.00 & -20.18 & -2.22  & 0.002 
\enddata

\tablecomments{The parameters $\phi*$, $M*$, and $\alpha$ represent the three parameters of the Schechter UV LF taken from the different papers.}

\label{tab:assumedLFs}
\end{deluxetable}

\section{Summary}
\label{sec:summary}

In this paper we present \textit{HST} slitless grism spectra for a uniquely bright $z>10$ galaxy candidate, which we previously identified in the GOODS-North field, GN-z11.
Our 2D data show clear flux longward of $\sim1.47~\mu$m exactly along the trace of the target galaxy and  zero flux at shorter wavelengths, thanks to our comprehensive and accurate treatment of contamination by neighboring galaxies. The interpretation that we indeed detect the continuum flux from GN-z11 is supported by the morphology of the spectrum, the fact that the counts fall off exactly where the sensitivity of the G141 grism drops, as well as the consistency of the observed counts with the $H$-band magnitude of GN-z11 (see e.g. Fig \ref{fig:2Dgrism}).

The grism spectrum, combined with the photometric constraints, allows us to exclude plausible low-redshift SEDs for GN-z11 at high confidence. In particular, we can invalidate a low redshift SED of an extreme line emitter galaxy at $z\sim2$ (see section \ref{sec:results} and Fig \ref{fig:ContamSpectra}).
Instead, the grism spectrum is completely consistent with a very high-redshift solution at $z_\mathrm{grism}=11.09^{+0.08}_{-0.12}$ (see Figures \ref{fig:2Dgrism} and \ref{fig:currentGrism}). This indicates that this galaxy lies at only $\sim400$ Myr after the Big Bang, extending the previous redshift record by $\sim150$ Myr.

GN-z11 is not only the most distant spectroscopically measured source, but is likely even more distant than all
other high-redshift candidates with photometric redshifts, including MACS0647-JD at $z_\mathrm{phot} = 10.7^{+0.6}_{-0.4}$ \citep[][]{Coe13}.
Additionally, GN-z11 is surprisingly bright, being among the brightest of any galaxies currently identified at $z>6$ (see Figure \ref{fig:ZspecMuv}). An SED fit to the photometry indicates that GN-z11 has built up a relatively large stellar mass ($\sim10^9~M_\odot$) for a galaxy at such an early time.
This unexpectedly luminous galaxy may challenge our current understanding of early galaxy build-up.
While the UV luminosity function of galaxies is not yet very accurately measured at $z>8$, the expected number density of such bright galaxies at $z\sim11$ is extremely small in most model estimates ($<0.3$ deg$^{-2}$). The fact that such a galaxy is found in only 0.2 deg$^2$ of the joint ACS and WFC3/IR data from the CANDELS survey is therefore somewhat surprising. Future surveys of at least several deg$^2$ will be required to accurately determine the number densities of such bright galaxies at $z>8$ and to characterize the bright end of the UV luminosity function. In particular, the planned high-latitude survey with WFIRST reaching to $H>26$ mag is expected to find a significant number of such bright sources.

The spectroscopic measurement of GN-z11 as a high-redshift source proves that massive galaxies of a billion solar masses already existed at less than 500 Myr after the Big Bang and that galaxy build-up was well underway at $z>10$. This is also promising news for future observations with the upcoming \textit{James Webb Space Telescope} (\textit{JWST}), which will be able to find galaxies at even earlier times. 
While challenging with \textit{HST} now, \textit{JWST}/NIRSPEC observations will be extremely efficient at confirming and measuring redshifts for all the current bright $z\gtrsim9$ galaxy candidates. JWST will push to much earlier times and also result in much larger samples of spectroscopically-confirmed sources within the first 500 Myr of cosmic time, particularly at $z\lesssim12$ if WFIRST is launched early enough to overlap with JWST. Until JWST, however, GN-z11 is quite likely to remain the most distant confirmed source.

\vspace*{0.5cm}

\acknowledgments{ We thank the anonymous referee for a helpful report which greatly improved this manuscript.	
The primary data for this work were obtained with the \textit{Hubble Space Telescope} operated by AURA, Inc. for NASA under contract NAS5-26555. Furthermore, this work is based in part on observations made with the \textit{Spitzer Space Telescope}, which is operated by the Jet Propulsion Laboratory, California Institute of Technology under a contract with NASA. This work is supported by NASA grant HST-GO-13871. }

Facilities: \facility{HST (ACS, WFC3), Spitzer (IRAC)}.

\bibliography{MasterBiblio}
\bibliographystyle{apj}

\appendix

\section{Consistency Checks of Grism Stacking and Extractions}
\label{app:grism}

\subsection{Median Stacking}

\begin{figure}[tb]
    \centering
        \includegraphics[width=0.65\linewidth]{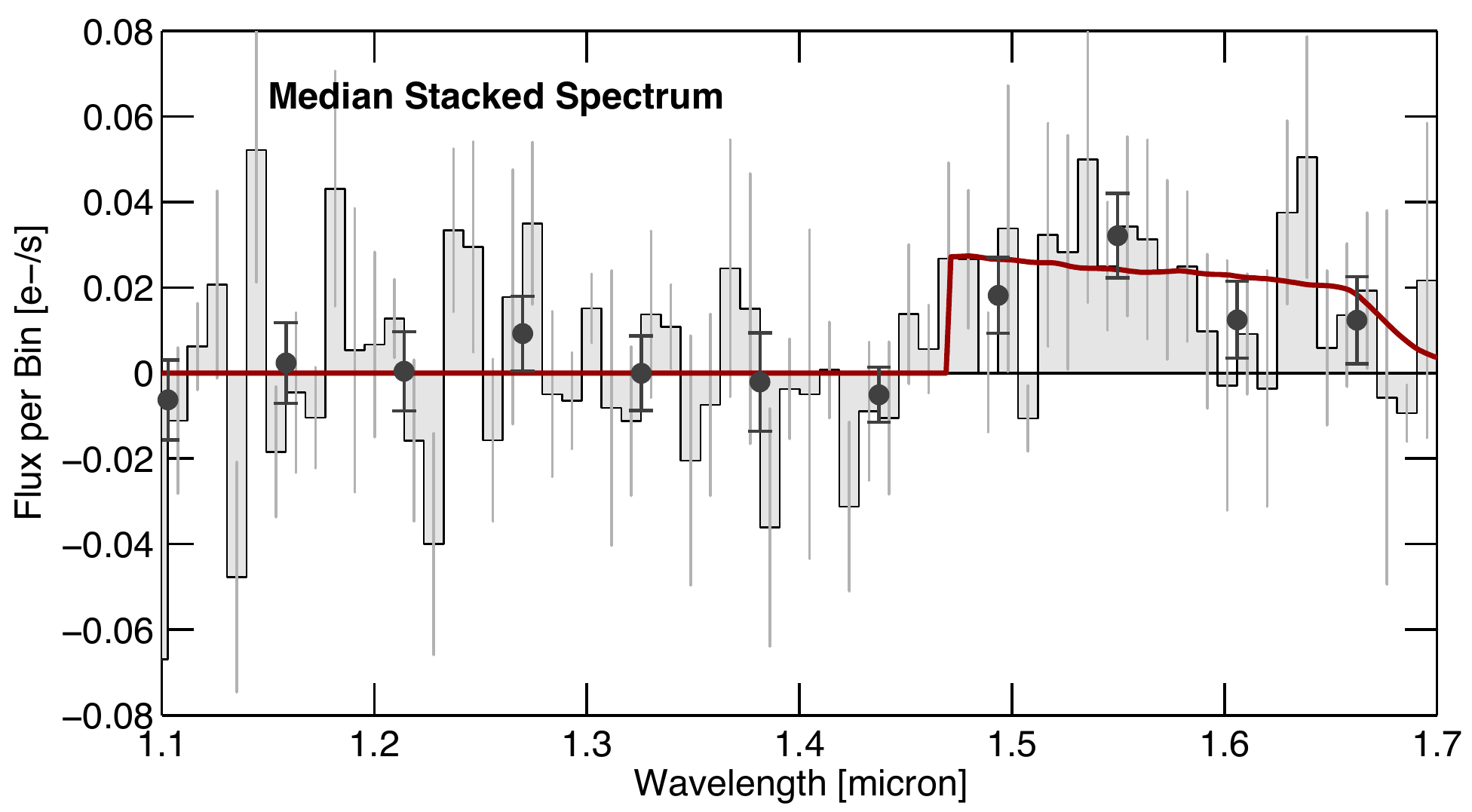}
    \caption{The 1D spectrum based on a simple median stack of the data from our 6 individual visits, rather than an optimal weighted stack. The gray histogram with errorbars shows the median spectrum rebinned to $\sim93$~\AA\ wide spectral bins. The median stacked flux still shows the spectral break at $\sim1.47$~\micron\ and is consistent with the expected count rate of an $H_{160} = 26$ mag source at $z_\mathrm{grism}=11.09$, which is shown by the red line. Black points with errorbars show the same 1D spectrum rebinned to 560 \AA. Note that this simple median stack does not optimally account for pixels affected by contamination and is overall noisier than the spectrum used in our main analysis.
 }
 \label{fig:medianStack}
 \end{figure}

In the main part of our analysis, we use a weighted sum to combine the 2D grism data from the 6 different visits to the final 12-orbit data. In particular, our weight includes a term to down-weight pixels which are affected by contamination (see Eq. \ref{eq:1}). To  ensure the detected signal is not the result of our stacking procedure, we additionally computed a simple median-stacked spectrum. This is shown in Figure \ref{fig:medianStack}. The median stack clearly still shows the continuum break, albeit it is somewhat noisier overall, since it does not include any contamination-based weighting. Nevertheless, the median flux is consistent with the expected count rate for an $H_{160}=26$ mag source at $z_\mathrm{grism}=11.09$.

\subsection{Extractions Across the Trace}

\begin{figure}[tb]
    \centering
        \includegraphics[width=0.5\linewidth]{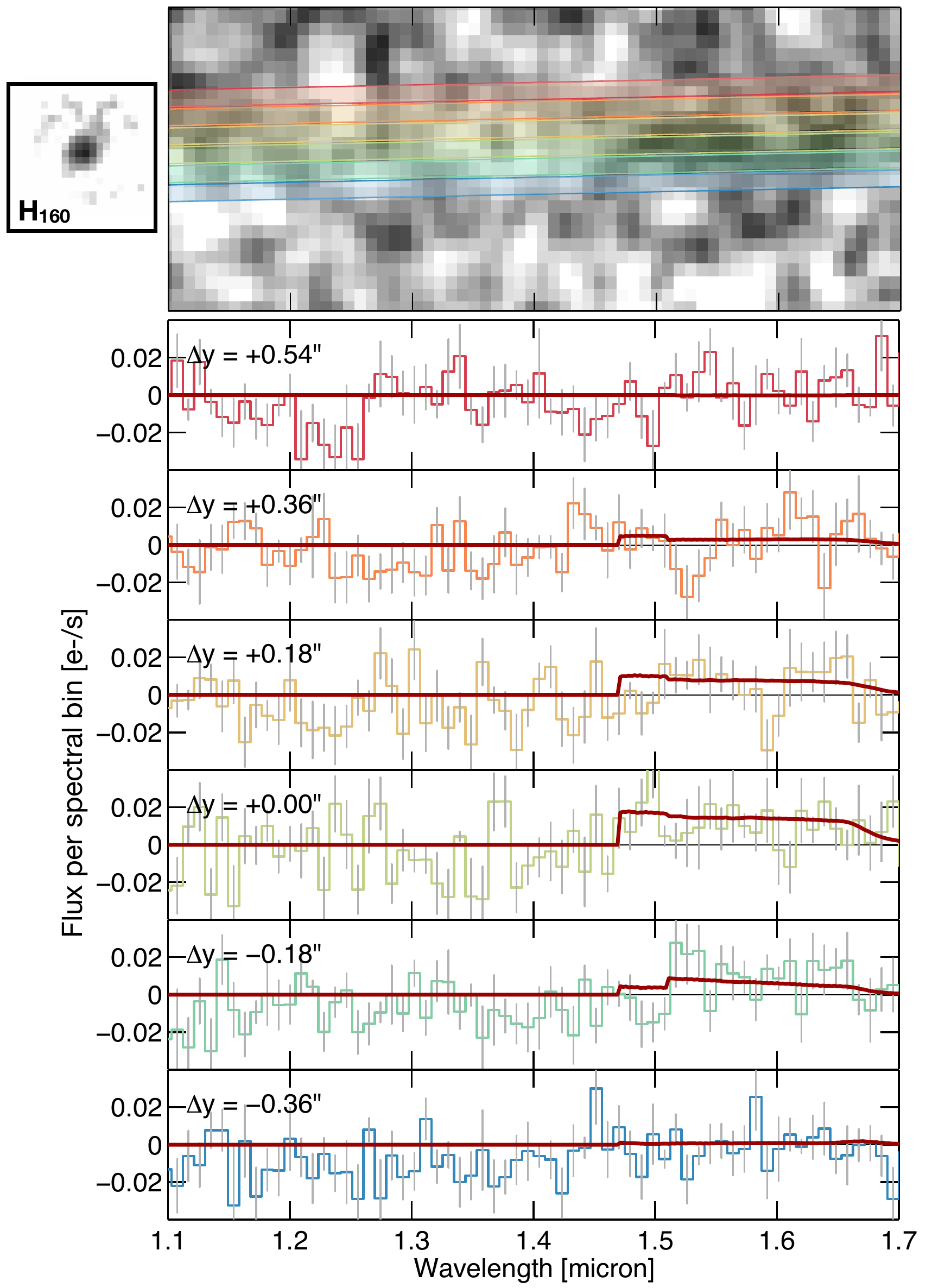}
    \caption{ 
1D spectra extracted at different offsets from the peak of the trace of GN-z11 (lower panels). The colored lines with errorbars show a simple sum of the flux over 0\farcs18 (3 pixels) in the spatial direction, offset in 0\farcs18 steps relative to each other. The regions over which the spectra are extracted are indicated in the top panel showing the 2D spectrum. The solid red line represents the expected 1D flux at each location based on our morphological model of GN-z11 with $H_{160}=26$ mag and at a redshift of $z_\mathrm{grism}=11.09$. (The discontinuities arise because the trace is slightly tilted.) Note that the asymmetric profile of GN-z11 extending over 0\farcs6 (see $H_{160}$-band stamp in the upper left) results in an asymmetric flux distribution relative to the peak of the trace. The negative dip at $\sim1.6$~\micron\ is seen most pronounced at $+$0\farcs18 above the peak of the trace of GN-z11.
 }
 \label{fig:offTrace}
 \end{figure}

The 1D spectrum in the main body of the paper is based on an optimal extraction taking into account the asymmetric morphology of GN-z11. Figure \ref{fig:offTrace} shows different 1D extractions as a function of position relative to the peak of the trace of GN-z11. These are simple sums extending over 0\farcs18 (i.e.\ 3 pixels) in the spatial direction. These 1D spectra show consistency with the expected count rate from our $H$-band morphological model of GN-z11. The figure also shows that the negative dip at $\sim1.6$~\micron\ stems from negative pixels slightly above the peak trace of GN-z11. We ensured that these negative pixels are not the result of any cosmic rays or inaccurate persistence or contamination subtraction. The dip extending over 3 rebinned pixels is consistent with simple Gaussian fluctuations based on our noise model.

\subsection{Data Split By Epoch}

\begin{figure*}[tb]
  \begin{center}
  \includegraphics[width=.99\linewidth]{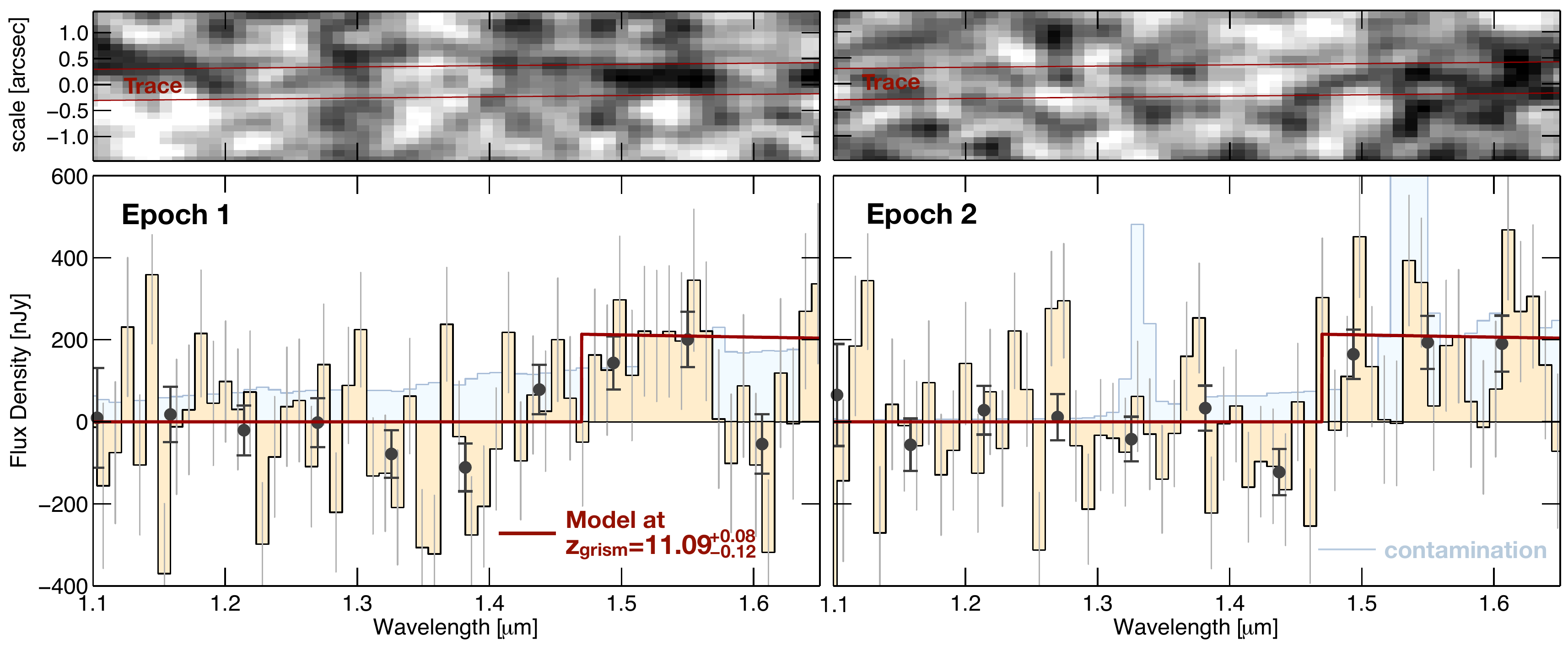}
  \end{center}
    \caption{ The 2D (top) and 1D (bottom) grism spectra of GN-z11 split in the two independent epochs of data from our program (epoch 1 left panels; epoch 2 right panels). Both of these have an exposure time of 6 orbits each, but their neighbor contamination is very different due to the different orientations (see Figure \ref{fig:ContamModel}). The blue lines show the contamination level which was subtracted from the 1D spectra.  As in Fig \ref{fig:currentGrism}, the top panel has been slightly smoothed for clarity and the lower panels show the 1D flux density binned to one resolution element of the G141 grism (93~\AA). As expected for a 6 orbit exposure, the continuum S/N is $\lesssim1$ at this resolution. The black dots with errorbars are the binned spectrum to 560~\AA. The continuum break is detected in both epochs separately.
   }
   \label{fig:SpecByEpoch}
\end{figure*}

We also tested whether a break is seen when further splitting up the data into our two independent epochs. The S/N of the continuum detection in the final 12-orbit stack shown in the main body of the paper is already relatively low, but this split-data test is a good cross-check on the viability and consistency of the result. These data of the individual epoch each consist of 6 orbits but at two different orientations, resulting in different contamination levels as a function of wavelength. Figure \ref{fig:SpecByEpoch} shows both the 2D and the corresponding 1D spectra for both epochs. When rebinned to 560 \AA\ a continuum break is seen in both epochs separately, consistent with our best-fit redshift $z_\mathrm{grism}=11.09$. However, the S/N at 93 \AA\ resolution in these spectra is small and the detailed differences should not be over-interpreted.

\begin{figure}[tb]
    \centering
        \includegraphics[width=0.5\linewidth]{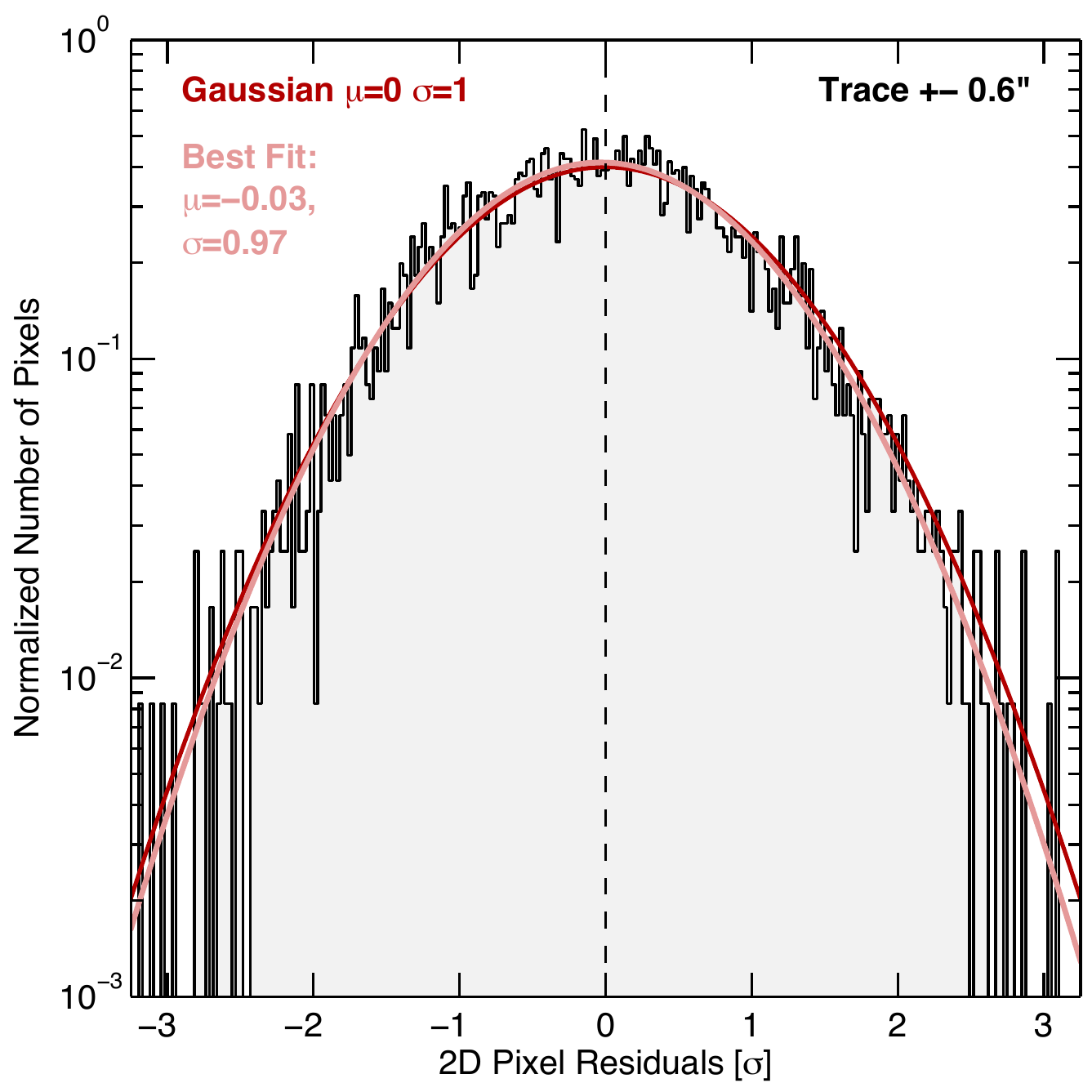}
    \caption{Histogram of pixel residuals within $\pm$0\farcs6 of the trace in the 2D spectrum after subtracting out the best-fit model of GN-z11. The salmon colored line is the best-fit Gaussian distribution, which is in excellent agreement with the expectation of a Gaussian centered at $\mu=0$ with $\sigma=1$ (dark red line). This demonstrates that our pixel noise estimates are accurate and that our extracted spectra are not affected by systematic uncertainties from neighbor modeling.
 }
 \label{fig:residualHist}
 \end{figure}

While some pixels off the trace in Fig. \ref{fig:SpecByEpoch} show residual flux in the epoch-split data, these pixels are heavily contaminated by neighboring galaxies and are downweighted in the final stack. In Fig. \ref{fig:residualHist} we show the final stacked pixel flux distribution within 0\farcs6 of the trace, which is perfectly consistent with a Gaussian. This demonstrates that our neighbor subtraction model and our pixel RMS estimates are both accurate.

\section{Keck MOSFIRE Spectroscopy}

Before the \textit{HST} grism spectra, GN-z11 was also observed with ground-based near-infrared spectra. We used Keck/MOSFIRE on 2014 April 25 to obtain $J$, $H$, and $K$-band coverage of GN-z11 (as well as lower redshift filler targets). The exposure times were 2.8 hrs in $J$, 1.0 hr in $H$, and 1.0 hr in $K$. The main purpose of these spectra was to rule out strong emission line contamination. With a seeing ranging from 0.8$-$1.4$\arcsec$ and some cirrus clouds throughout the night, these spectra are not extremely constraining. However, the higher spectral resolution compared to the WFC3/IR grism allows us to search for narrower emission lines. None were found. In between sky lines, our spectra are sensitive to emission line fluxes of $\sim2-4\times10^{-17}$ erg\,s$^{-1}$\,cm$^2$ (5$\sigma$). These observations therefore already provided some evidence against strong emission line contamination before the acquisition of the WFC3/IR grism spectra.

\section{Estimating Stellar Population Properties}
Spectral energy distribution fitting to the photometry was used to estimate several stellar population properties of GN-z11 listed in Table \ref{tab:specsummary}. This was done using the code ZEBRA+ \citep{Oesch10c}, an extension of the photometric redshift code ZEBRA \citep{Feldmann06}. The stellar population templates were based on standard libraries \citep{Bruzual03}, however, nebular continuum and line emission were added self-consistently assuming all ionizing photons are transformed to nebular emission. The allowed star-formation histories were based on standard exponentially declining models with parameters $\tau=10^{8},~10^{9}$, and constant star-formation, with stellar metallicities of $Z=0.05-0.5Z_\odot$, and a Chabrier initial mass function. The ages of the models ranged from $10^{6}$~yr to the age of the universe at the given redshift. Dust extinction was allowed in the fit using a standard starburst dust model \citep{Calzetti00}. 
The resulting stellar mass derived for GN-z11 is $\log M/M_\odot=9.0$ with formal uncertainties $\Delta \log M=0.4$ as derived from the $\chi^2$ values of all SEDs in the library. We note, however, that our longest wavelength filter, IRAC channel 2, only partially covers the rest-frame optical for GN-z11 at $\lambda_\mathrm{rf}>4000$~\AA, which is why we only report an approximate stellar mass in the main body of the text. Other physical parameters are listed in Table \ref{tab:specsummary}.

\end{document}